\documentclass[]{IEEEtran}
\usepackage{epsfig}
\begin{document}

\title{Redundancy and Robustness of the AS--level\\ Internet topology and its models}

\author{Shi Zhou and Ra\'ul J. Mondrag\'on
\thanks{Manuscript received on 19 August 2003, published on 22 January 2004 in IEE Electronic Letters vol. 40, no. 2, pp. 151-15.
This work is supported by the U.K. Engineering and Physical
Sciences Research Council (EPSRC) under Grant GR-R30136-01.}
\thanks{S. Zhou and R. J. Mondrag\'on are with the Department of Electronic Engineering,
Queen Mary, University of London, Mile End Road, London, E1 4NS,
United Kingdom (e-mail: shi.zhou@elec.qmul.ac.uk;
r.j.mondragon@elec.qmul.ac.uk). }}

\markboth{IEE Electronic Letters}{IEE Electronic Letters}

\maketitle

{\bf Abstract}: A comparison between the topological properties of
the measured Internet topology, at the autonomous system level (AS
graph), and the equivalent graphs generated by two different power
law topology generators is presented. Only one of the synthetic
generators reproduces the tier connectivity of the AS graph.\\

{\bf Introduction}: Simulation plays an important role in the
development of the Internet, as it can be used to compare and
analyze new network protocols. In these simulations it is crucial
that the topology generators capture the key topological
properties of the Internet. For example, Labovitz~{\sl et~al} [1]
showed that the topology of the Internet has a major impact on the
delayed BGP routing convergence.\\

If each autonomous system (AS) of the Internet is represented by a
node in a graph, Faloutsos~{\sl et~al} [2] discovered that the
link connectivity between these nodes follows the power law
$P(k)\propto k^{-\gamma}, \gamma\approx 2.2$, where $k$ is the
number of links a node has. A good model of the AS-level Internet
topology not only has to reproduce the power law link connectivity
but also the connectivity of the core of the network. Tier 1 of
the AS graph is the core of the network which consists of a set of
nodes which are very rich in links and are densely interconnected
with each other, we called this set of nodes the rich--club [3].
This letter shows that a power law topology generator without a
rich--club could under--estimate network redundancy and
over--estimate network robustness of the Internet.\\

\begin{table}[htb]\caption{Network properties}\centering
\renewcommand{\tabcolsep}{0.4pc} 
\renewcommand{\arraystretch}{1.2} 
\begin{tabular}{c c c c }
\hline
~               &   AS graph    &   IG graph    &   FBA graph\\
\hline
Number of nodes, $N$             &   11122       &   11122       &   11122\\
Number of links, $L$             &   30054       &   33349       &   33349\\
Power--law exponent, $\gamma$             &   2.2       &   2.22       &   2.255\\
Maximum $k$ & 2839       &    842        & 1793\\
Maximum~$K_{t}$ &    7482      &     4962     &  1191 \\
Average~$K_{t}$ &    12.7      &      10.0    &  0.6 \\
\hline
\end{tabular}\\[2pt]
\end{table}

{\bf Methodology}: We compared the traceroute AS graph [4]
measured on the 1st of April 2002 against the synthetic networks
generated by the Fitness Barab\'asi--Albert (FBA) model [5] and
Interactive Growth (IG) model [6]. The two models create networks
using a node growth mechanism, where a new node attaches to other
nodes and prefers to attach itself to nodes that have large
numbers of links. The FBA model is an example of a generator where
the preferential attachment is controlled by a fitness parameter.
This parameter adjust the node's ability of acquiring connections
with other nodes. The IG model is an example of a generator where
a new node attaches itself to other nodes and also creates new
links between nodes that already exist on the network. As shown in
table 1, these two topology models generate networks that have
similar sizes and power law degree distributions as the AS
graph.\\

\begin{figure}[htb]
\centerline{\psfig{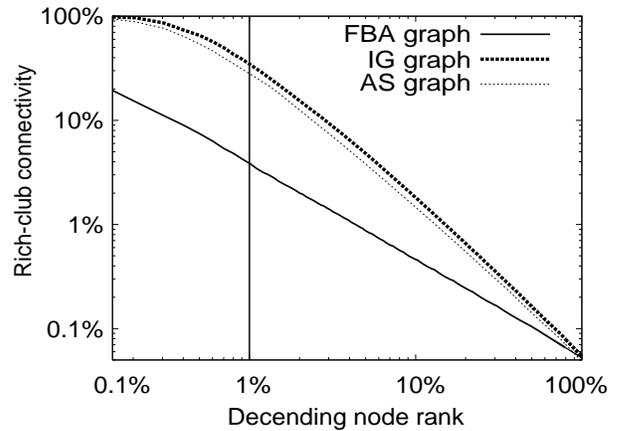}}\caption{Rich--club
connectivity.}
\end{figure}

Each node in a network is sorted in decreasing number of the links
that the node contains. The node rank $r$ is the position that a
node has on this ordered list. The position is normalized by the
total number of nodes $N$. The rich--club are the nodes with rank
less than $r$ (e.g. 1\%). The core connectivity of the network
(tier 1) is measured using the rich--club connectivity, which is
the ratio of the actual number of links between the members of the
rich--club to the maximum possible number of links (fully
connected). Figure 1 shows that the IG graph closely matches the
AS graph's rich--club connectivity. For example, the 1\% best
connected nodes have 35\% of the maximum possible number of links,
whereas in the FBA graph they only have 4\% of the maximum
possible number of links.

In a network a circuit of length three is called a triangle. The
number of alternative routes in a network increases with the
number of triangles. The triangle coefficient $K_t$ of a node is
defined as the number of triangles that share the node. Figure 2
and table 1 show that the AS graph and the IG graph have
significantly more triangles than the FBA graph. Hence the FBA
model produces networks that are less flexible to traffic routing
and have a lower degree of network redundancy.\\

\begin{figure}[htb]
\centerline{\psfig{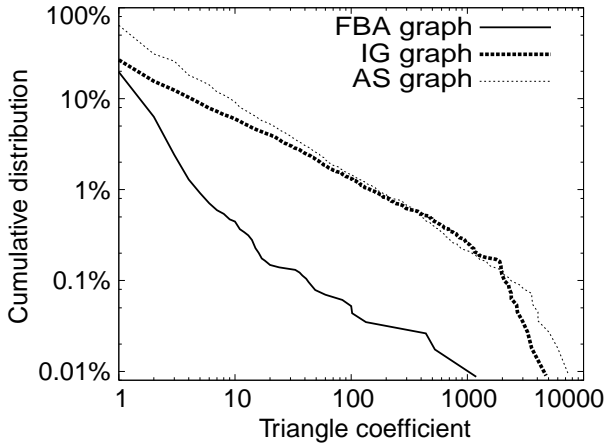}}\caption{Cumulative
distribution of triangle coefficient.}
\end{figure}

\begin{figure}[htb]
\centerline{\psfig{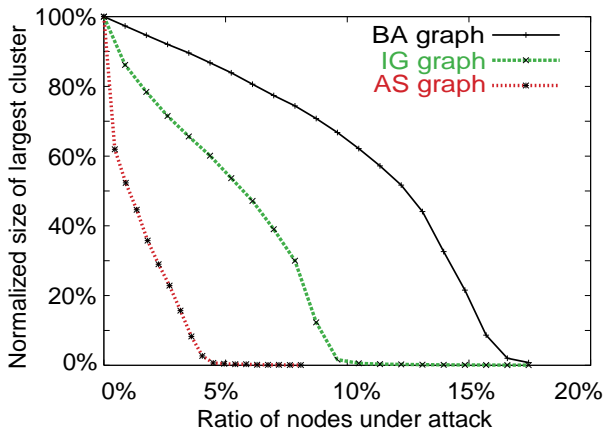}}\caption{Network
robustness under node attack.}
\end{figure}

The removal of the best connected nodes from a network is called
node attack. This resembles the Internet scenario where a node, AS
or router, is out of service due to severe traffic congestion or
infection by a malicious virus. Figure 3 shows that the AS graph
and the IG graph are extremely fragile under node attack. The
removal of only a few of the AS graph's richest nodes can break
down the network integrity. By comparison, the FBA graph is not so
vulnerable and shows higher degree of resilience to node attack.\\

{\bf Conclusion}: The AS graph have a densely interconnected core
structure, which plays a dominant role in the network. The
rich--club acts as a super traffic hub providing a large selection
of shortcuts. Realistic models of the Internet should reproduce
this core structure to avoid under--estimating the network
redundancy and over--estimating the network robustness.\\

\end{document}